\documentclass[a4paper]{jpconf}
\usepackage{graphicx}
\usepackage{subfig}
\usepackage[export]{adjustbox}
\usepackage{float} 
\usepackage{xurl} 
\usepackage[margin=1in]{geometry}
\usepackage{hyperref}
\hypersetup{
    colorlinks=true,
    citecolor=blue,
    linkcolor=blue,
    filecolor=blue,      
    urlcolor=blue,
    }

\begin{document}
\title{Facilitating Non-HEP Career Transition}
\author{Sudhir Malik$^1$, Aneliya Karadzhinova-Ferrer$^{2,3}$, Julie Hogan$^4$, Rachel Bray$^5$, Rami Kamalieddin$^6$, Kevin Flood$^7$, Amr El-Zant$^8$, Guillermo Fidalgo$^1$, David Bruhwiler$^9$, Matt Bellis$^{10}$ }

\address{$^1$Physics Department, University of Puerto Rico Mayagüez, PR 00682, USA}
\address{$^2$Helsinki Institute of Physics, Finland}
\address{$^3$Lappeenranta-Lahti University of Technology, Finland}
\address{$^4$Department of Physics and Engineering, 3900 Bethel Dr, St Paul, MN 55112, USA}
\address{$^5$CERN, Switzerland}
\address{$^6$AllianceBernstein Investment Management,New York, NY }
\address{$^7$Physics Department, University of Hawaii, HI 96822, U.S.A; Nalu Scientific LLC}
\address{$^8$Centre for Theoretical Physics, The British University in Egypt, Cairo 11837, Egypt}
\address{$^9$RadiaSoft LLC, Boulder, Colorado, CO 80301, USA}
\address{$^{10}$Siena College, 515 Loudon Road, Loudonville, NY 12211, USA}
\ead{sudhir.malik@upr.edu}

\begin{abstract}
About two-third of Physics PhDs establish careers outside of academia and the national laboratories in areas like Software, Instrumentation, Data Science, Finance, Healthcare, Journalism, Public Policy and Non-Governmental Organization. Skills and knowledge developed during HEPA (High Energy Physics and Astrophysics) research as an undergraduate, graduate or a postdoc level (collectively called early career) have been long sought after in industry. These skills are complex problem solving abilities, software programming, data analysis, math, statistics and scientific writing, to name a few. Given that a vast majority transition to the industry jobs, existing paths for such transition should be strengthened and new ways of facilitating it be identified and developed. A strong engagement between HEPA and its alumni would be a pre-requisite for this. It might also lead to creative ways to reverse the "brain drain" by encouraging alumni to collaborate on HEPA research projects or possibly come back full time to research. We motivate and discuss below several actionable recommendations by which HEPA institutions as well as HEPA faculty mentors can strengthen both ability to identify non-HEP career opportunities for students and post-docs as well as help more fully develop skills such as effective networking, resume building, project management, risk assessment, budget planning, to name a few.   This will help prepare early career HEPA scientists for successfully transitioning from academia to the diverse array of non-traditional careers available. HEPA alumni can play a pivotal role by engaging in this process.

\end{abstract}

\section{Introduction}
HEPA research  gives undergraduate or  graduate (Masters or PhD) students or postdocs a wide array of scientific and technical skills. There are a variety of career paths available to physics degree holders beyond  academia, where these skills are in demand. The academic job market is such that faculty positions and scientists in universities and other institutions is very stable, meaning that not many new jobs are being created. According to AIP~\cite{nonhep_aip} report~\cite{nonhep_aipindustry} more than two-thirds of Physics PhD recipients accept permanent jobs in industry. Not only does industry provide the highest number of jobs~\cite{aip_afterPhD, phytoday_afterPhD,physicstoday_careers} but also the highest salary. Early career physicists may exit academia at different stages - post undergrad, masters or a PhD or postdoc or even later. While the time spent in doing PhD in HEPA are typical compared to the other fields, postdocs usually tend to spend 5-6 years, a bit on higher side compared to other disciplines,  before landing an academic job or leaving for industry.  Irrespective of stage of finding the first job, the evolution from academic preparation to the first steps in career may not be easy to predetermine. A smart planning, awareness of various possibilities in industry for jobs, accompanied with career networking with alumni and colleagues, can facilitate the process of industry job search at all stages. 

There are efforts in the above direction by AIP~\cite{nonhep_aipcareer} for all Physics disciplines. However, networking events for industry careers targeted for HEPA community were set off at CERN by the CMS Collaboration~\cite{nonhep_networkCMS2013,nonhep_networkCERN2013} a decade ago and later adopted by all CERN Experiments~\cite{nonhep_networkCERN2016,nonhep_networkCERN}. It eventually led to the creation of the CERN Alumni Network~\cite{nonhep_networkCERNalumni} and the CERN Career Fairs~\cite{nonhep_networkCERNcareer}. In the USA there are some efforts in that direction;  like those by the Fermilab Student Postdoc Association~\cite{nonhep_fspafermilab}. But clearly a more organised effort is required. The creation of targeted internships with national laboratory partners in the areas of Accelerator Technology, Computer and Information Science, Detector and Engineering Technology, Environmental Safety and Health and Radiation Therapies is one way to facilitate the training process for a career in industry. However, many postdocs and students are not based at National labs where broad access to the above opportunities exists. Direct involvement from supervisors and mentors in planning the career of their mentees early on, based on the interest, is another important step towards success. A commensurate effort in the job search process is also needed. A centralised method of networking with HEPA alumni would strengthen the job search, skill matching and hiring process. However, there is a continuing need for new and innovative tools and portals to be developed and supported. A robust association and collaboration with HEPA alumni on research projects could also possibly lead to reversing of the “brain drain” from HEPA. Feedback based on a survey performed in the HEPA community, described below, suggests steps which can be taken to further develop pathways for efficient and effective transition for HEP scientists, at all levels,leading to a successful career in industry. The goal of this paper is twofold:  first to motivate the need for such improvements, and then to subsequently identify a variety of mechanisms with which to achieve the above goals.

\section{Survey findings/analysis, anecdotal experiences}

The Snowmass Early Career~\cite{nonhep_snowmassyoung} team prepared and conducted a survey between June 28 and August 15, 2021 for the entire High Energy Physics and Astrophysics (HEPA) community. The survey was designed to inform the survey team on opinions, experiences, and outlook regarding several topics; including careers, physics outlook, workplace culture, harassment, racism, visa policies, the impacts of COVID-19, and demographics. The Community Engagement Frontier (CEF) Working group (WG) on Career Pipeline and Development (CPD) strategically planned and structured  questions and contributed to this survey to collect feedback related to its own goals. Some of the results of the survey are discussed in the following sections. The CPD questions focused on feedback from HEPA community with respect to preparedness for jobs and career plans outside HEPA and the mentoring received. For HEPA alumini the questions centered around the diverse occupation in industry, use of skills learned in HEPA, resources in obtaining the job and possible collaborations. The feedback from the survey informs on the measures that HEPA community should take to prepare and facilitate the transition to industry jobs; as early planning, organised networking, and mentoring, and building better relationships with alumni for mutual benefit. The survey also tells us that most people try their best to stay in HEPA as long as they can. More than 50\% of the participants are aged below 35 years and participants who declared their nationalities were from 60 different countries. The survey findings presented here are those that are most relevant to this paper. The feedback from survey results is divided into two groups: from HEPA community and HEPA alumni, as discussed below.

\subsection{Survey results from HEPA community}
\begin{figure}[h]
\centering
\begin{minipage}[b]{0.5\textwidth}
    \includegraphics[width=0.955\linewidth]{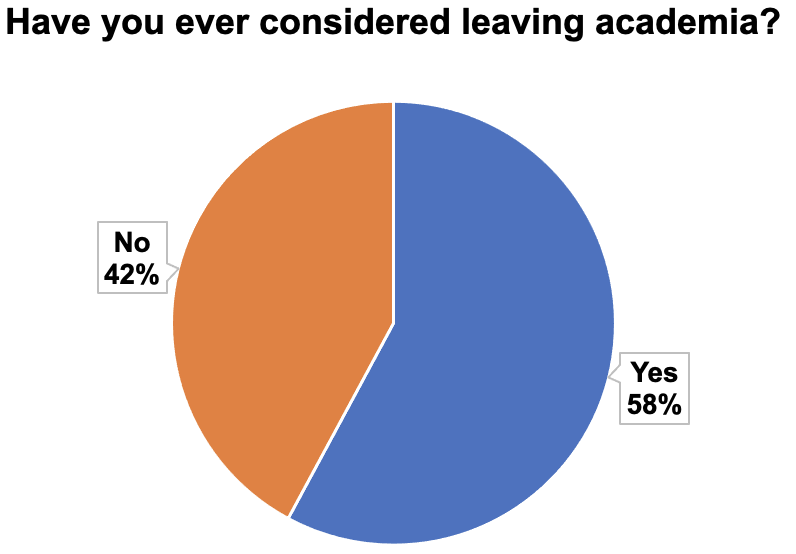}
 \caption{}
    \label{fig:nonhep_Q42}
    \end{minipage}
\end{figure}

\begin{figure}[h]
\centering
\includegraphics[width=35pc]{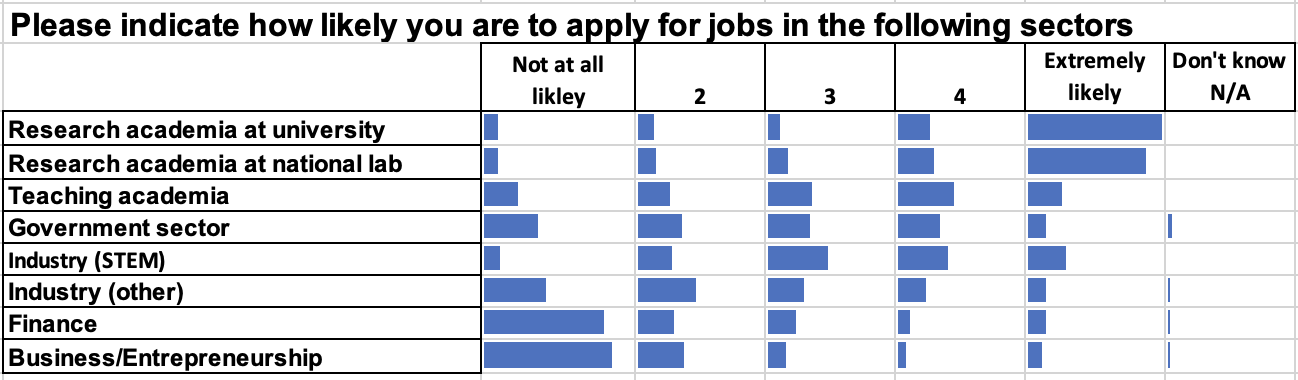}\hspace{5pc}%
\begin{minipage}[b]{28pc}\caption{\label{fig:nonhep_Q27}}
\end{minipage}
\end{figure}

The survey participants from HEPA community were from Undergraduates, Masters and PhD students, Postdocs, Engineers, Technicians, Teaching faculties, Tenure-track faculties, Tenured faculties and Scientists or Senior scientists from national lab and universities. Almost half of these survey participants are considering leaving academia (Figure~\ref{fig:nonhep_Q42}), making it essential to organise and facilitate this transition. While the first preference obviously is research jobs in academia (university faculty) or national labs (scientist), the preference for being in  business or being an entrepreneur is the least as shown in Figure~\ref{fig:nonhep_Q27}. 

\begin{figure}[h]
\centering
\includegraphics[width=35pc]{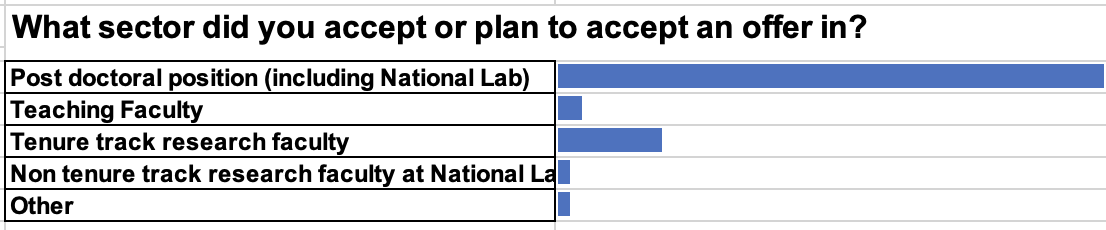}\hspace{5pc}%
\begin{minipage}[b]{28pc}\caption{\label{fig:nonhep_Q23}}
\end{minipage}
\end{figure}

\begin{figure}[h]
\centering
\includegraphics[width=30pc]{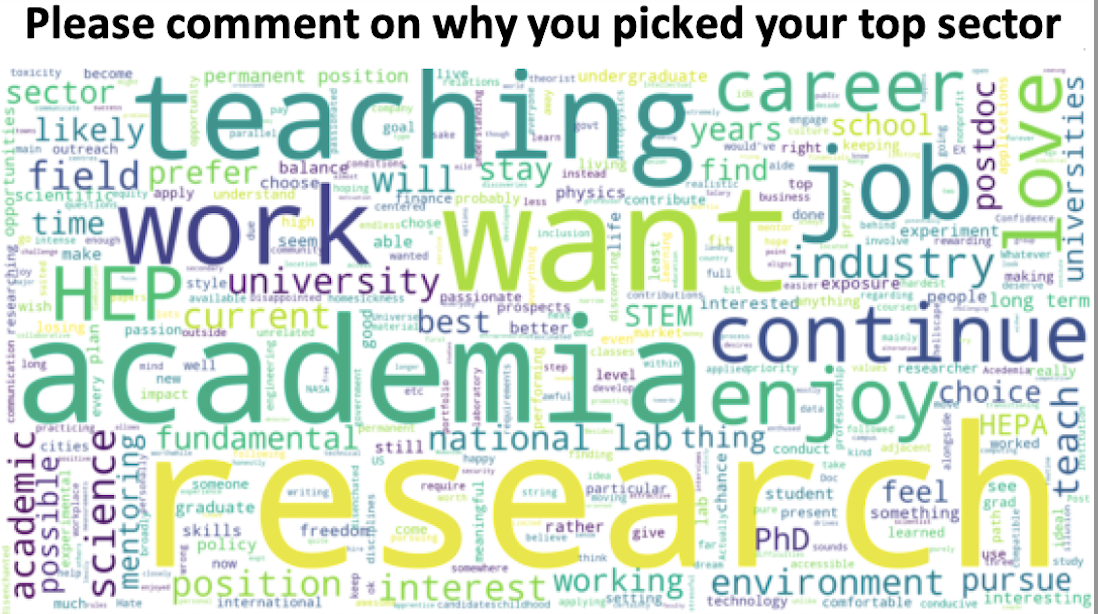}\hspace{5pc}%
\begin{minipage}[b]{28pc}\caption{\label{fig:nonhep_Q28}}
\end{minipage}
\end{figure}

\begin{figure}[h]
\centering
\begin{minipage}[b]{0.5\textwidth}
\includegraphics[width=0.955\linewidth]{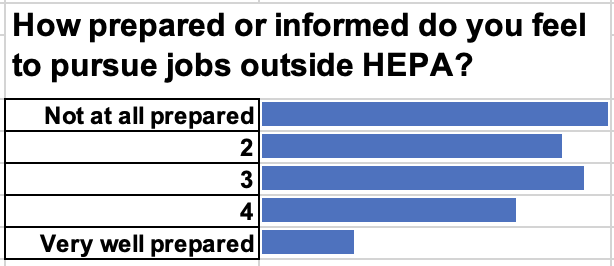}
\caption{}
\label{fig:nonhep_Q22}
\end{minipage}
\begin{minipage}[b]{0.4\textwidth}
\includegraphics[width=1.155\linewidth]{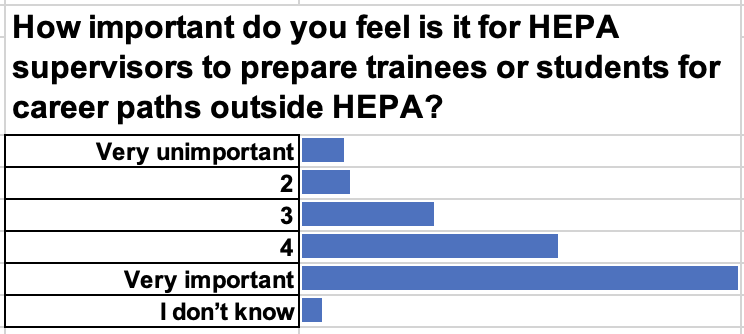}
\caption{}
\label{fig:nonhep_Q62}
\end{minipage}
\end{figure}

Most participants, being young, have accepted or plan to accept academic positions that can constitute an intermediate step in an academic career ladder; like going from PhD position to a postdoc, or from a postdoc to a  faculty position, as shown in Figure~\ref{fig:nonhep_Q23}, which again indicates a strong preference to stay in HEPA. The level of unpreparedness for non-HEPA jobs (industry jobs, teaching positions) is overwhelming, as indicated in Figure~\ref{fig:nonhep_Q22}. To continue to attract students to research and provide the strongest possible assurance of a secure and interesting career, we must as a community facilitate the process of transition to  positions in industry, as these form the bulk of the job market available to (and suitable for) HEPA-trained scientists. This would also provide a pipleline of talent to non-HEPA STEM areas, both within or outside academia. A free-text response to a question asking to pick a reason to apply for a future academic or industry job shows the top word used in Figure~\ref{fig:nonhep_Q28} as "research". This implies that career preference is to do research whether in acedemia or industry.

\begin{figure}[h]
\centering
\includegraphics[width=30pc]{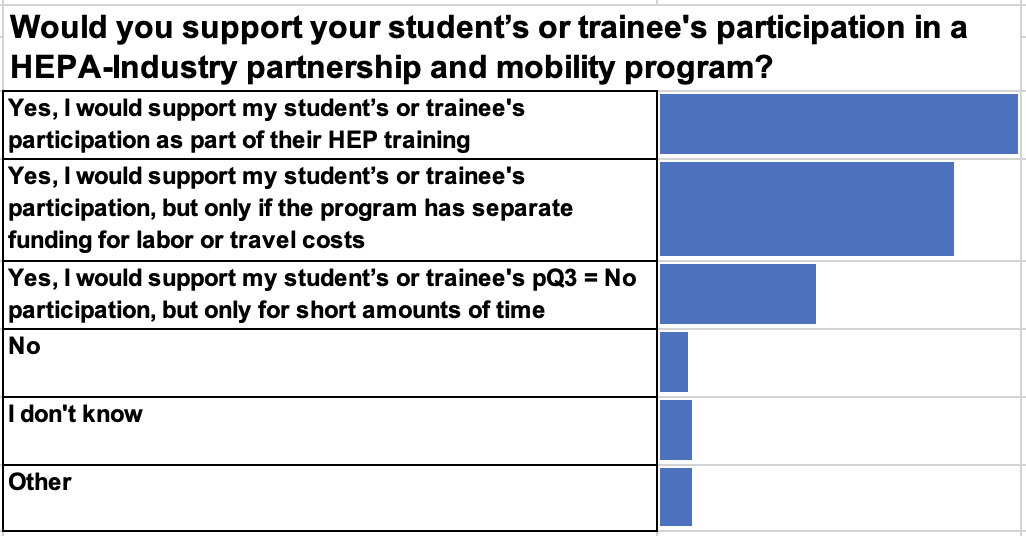}\hspace{5pc}%
\begin{minipage}[b]{28pc}\caption{\label{fig:nonhep_Q64}}
\end{minipage}
\end{figure}

\begin{figure}[h]
\centering
\includegraphics[width=35pc]{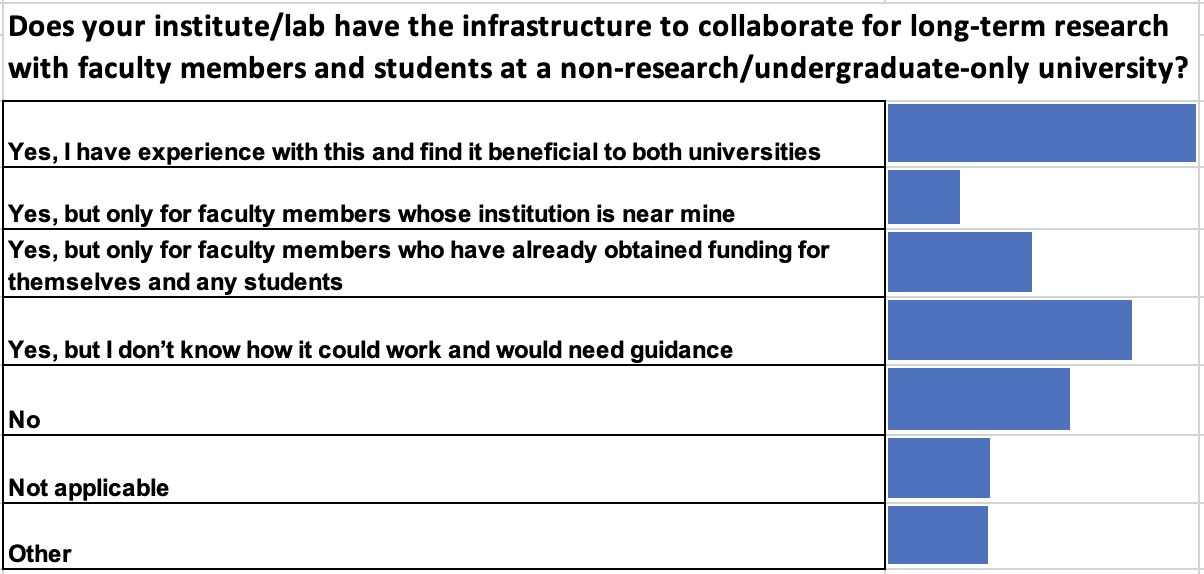}\hspace{5pc}%
\begin{minipage}[b]{28pc}\caption{\label{fig:nonhep_Q61}}
\end{minipage}
\end{figure}

While it is true that not enough mentoring and preparedness exists, there is a realisation among the HEPA community that it is important for supervisors and mentors to prepare their students for career paths in industry, as shown from the survey result in Figure~\ref{fig:nonhep_Q62}. Among supervisors, there is also strong support for their students and mentees to participate in HEPA-industry partnerships and mobility programs that would prepare them well for industry jobs, as shown in Figure~\ref{fig:nonhep_Q64}. However, it is also shown that funding for such partnership activities must come from a source separate from their core HEPA research support. Further interesting feedback is shown in Figure~\ref{fig:nonhep_Q61}, reflecting  strong  support for collaboration between research and non-research/undergraduate institutions. This is essential in building and broadening STEM opportunities that would provide a pipeline of scientific talent to the broader workforce. In general, physics skills are central to the landscape in industry,and offer routes to productive employment in varied and rewarding careers.


\subsection{Survey results from HEPA alumni}

\begin{figure}[ht]
\centering
\subfloat{
  \includegraphics[width=75mm]{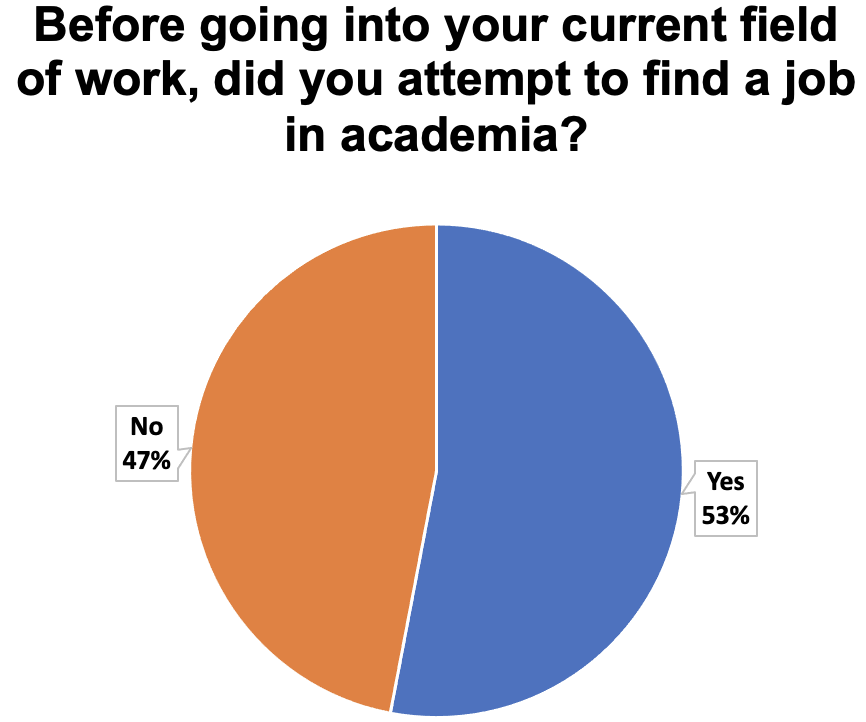}
}
\subfloat{
  \includegraphics[width=80mm]{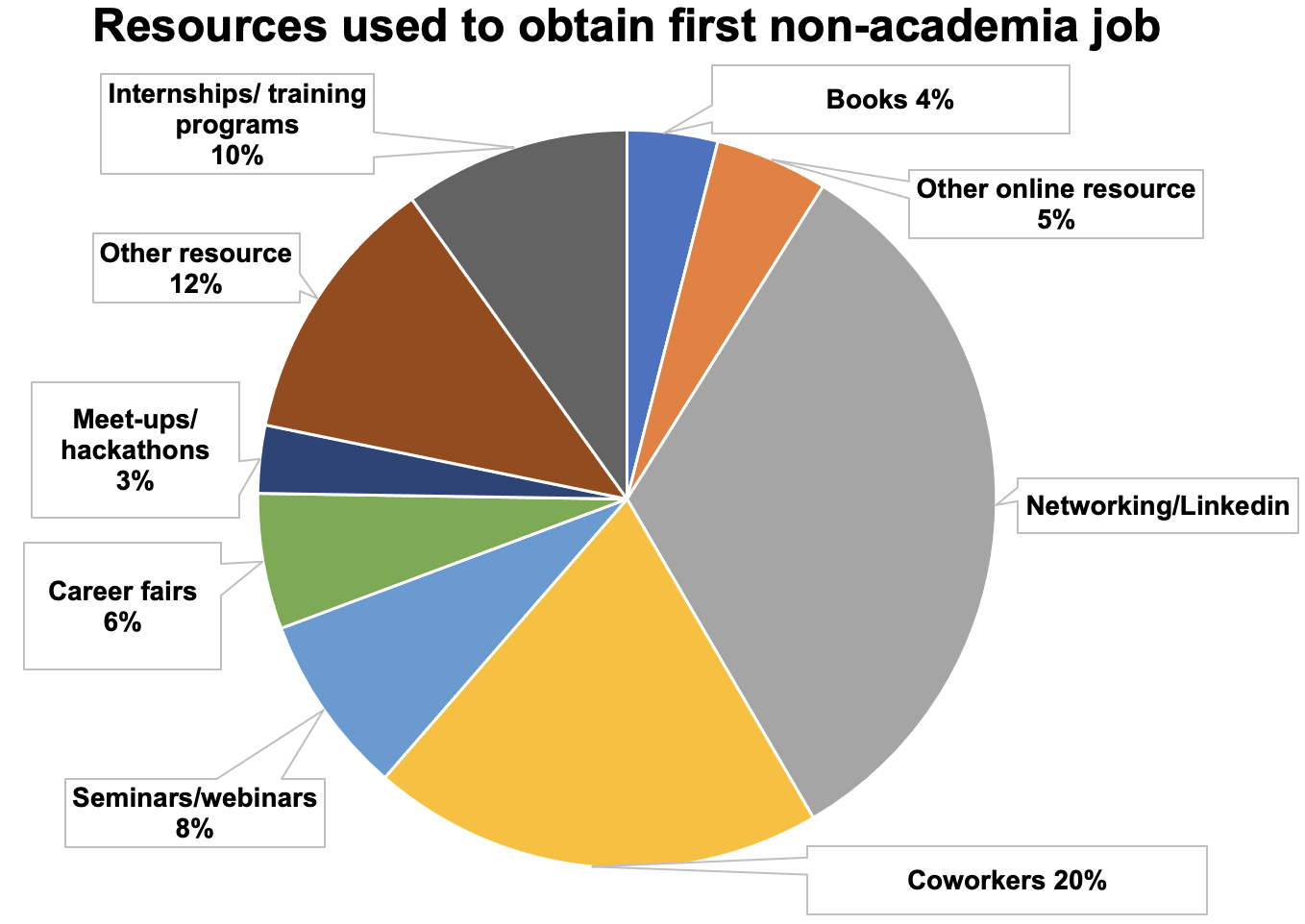}
}
\hspace{0mm}
\subfloat{
  \includegraphics[width=65mm]{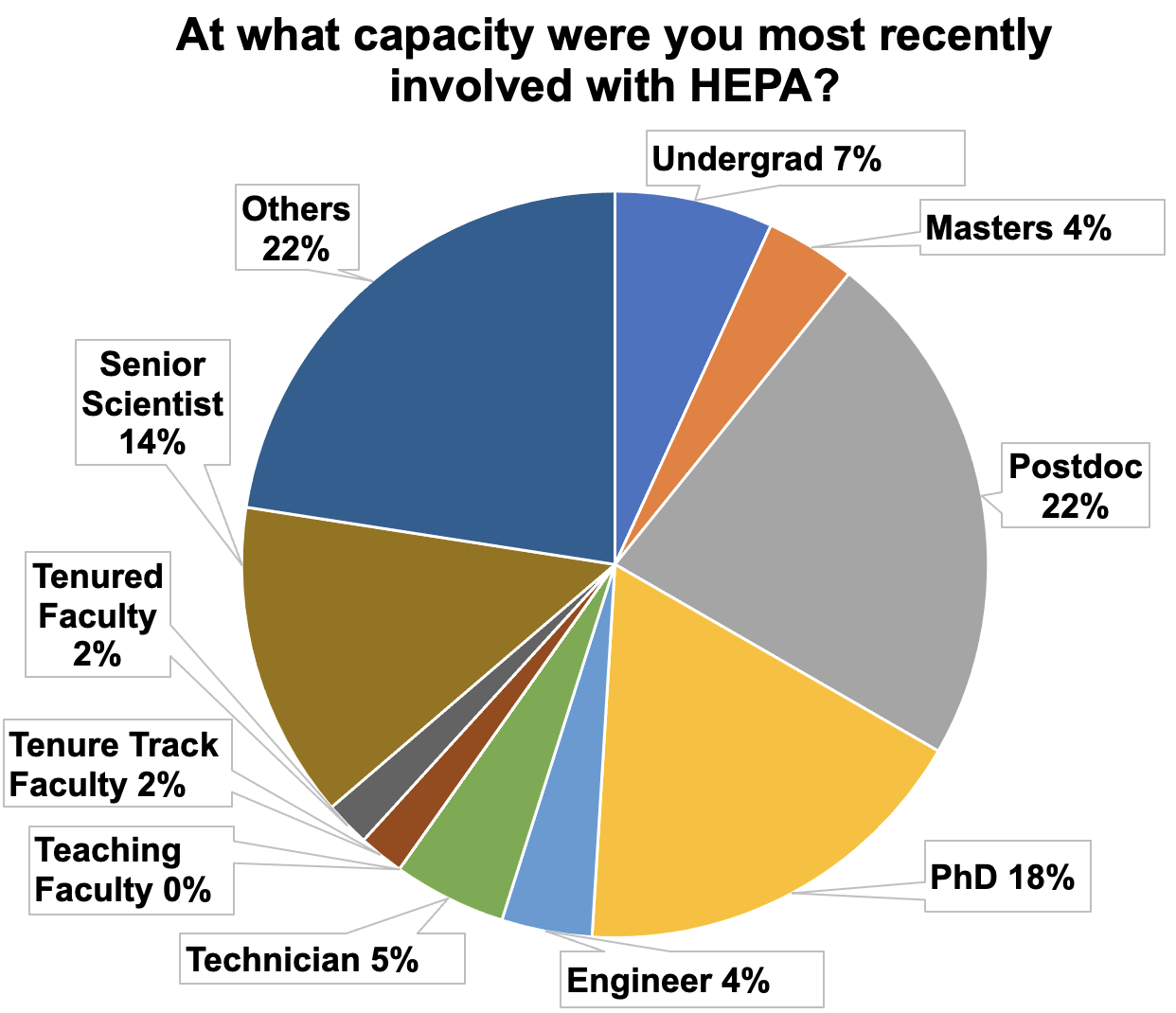}
}
\subfloat{
  \includegraphics[width=85mm]{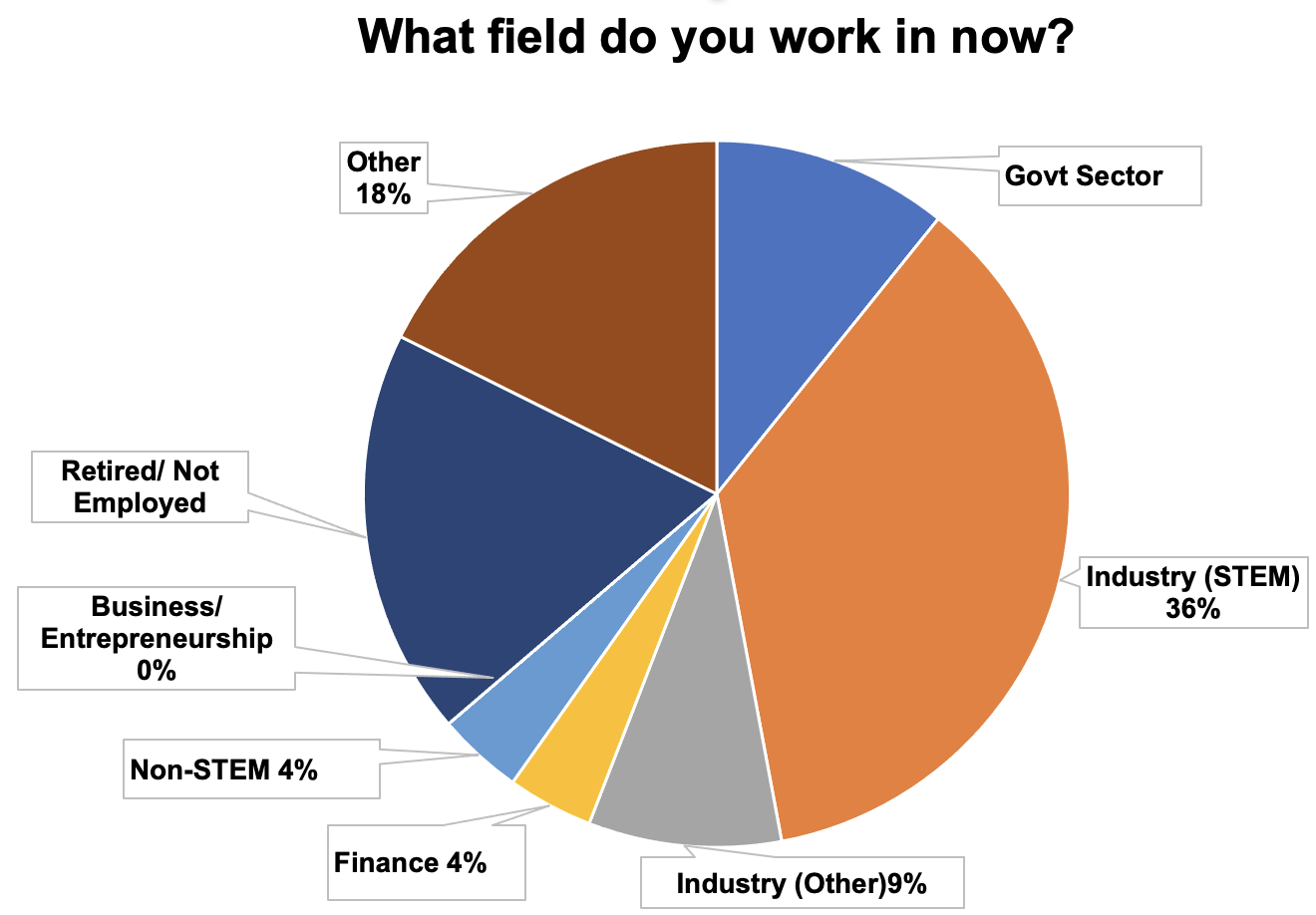}
}
\caption{\label{fig:nonhep_Q69Q70Q66Q68}}
\end{figure}

\begin{figure}[ht]
\centering
\includegraphics[width=35pc]{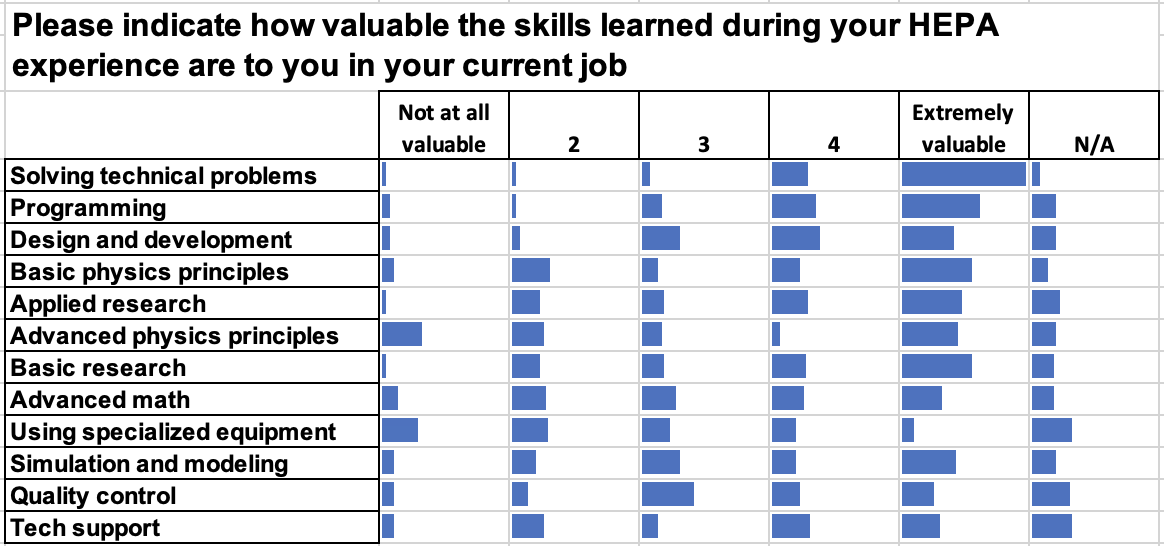}\hspace{5pc}%
\begin{minipage}[b]{28pc}\caption{\label{fig:nonHEP_Q71}}
\end{minipage}
\end{figure}

\begin{figure}[ht]
\centering
\includegraphics[width=35pc]{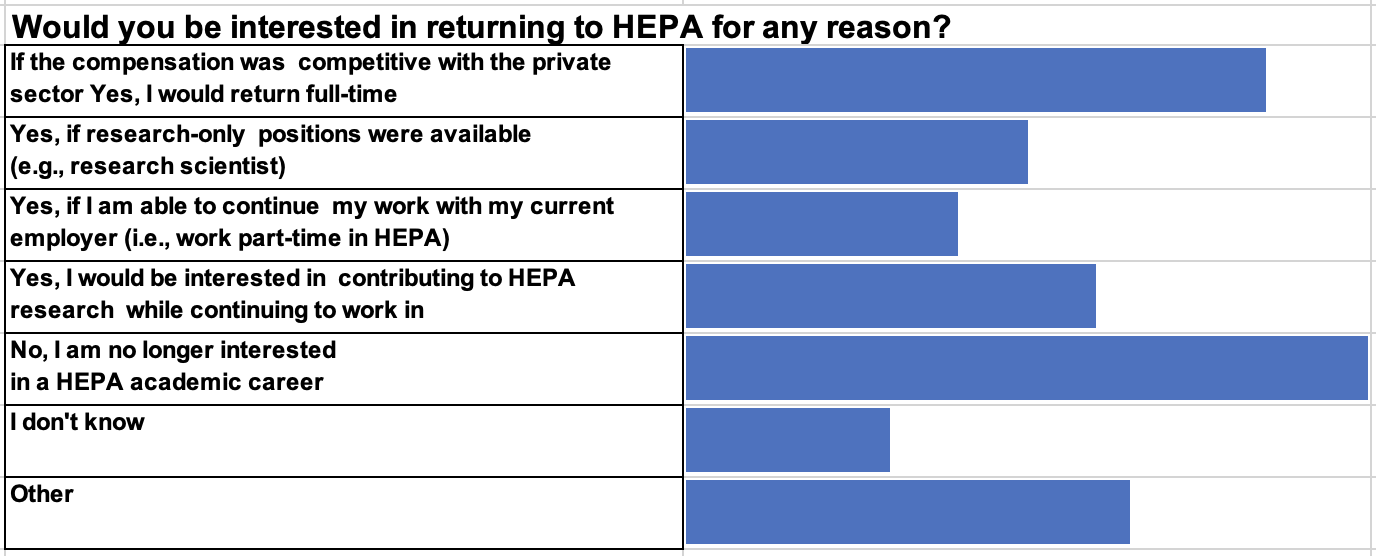}\hspace{5pc}%
\begin{minipage}[b]{28pc}\caption{\label{fig:nonhep_Q72}}
\end{minipage}
\end{figure}

\begin{figure}[h]
\centering
\begin{minipage}[b]{28pc}
\includegraphics[width=1.0\linewidth]{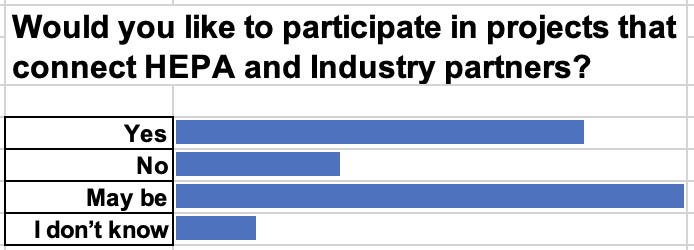}
\caption{}
\label{fig:nonhep_Q33}
\end{minipage}
\end{figure}

\begin{figure}[h]
\centering
\begin{minipage}[b]{28pc}
\includegraphics[width=25pc]{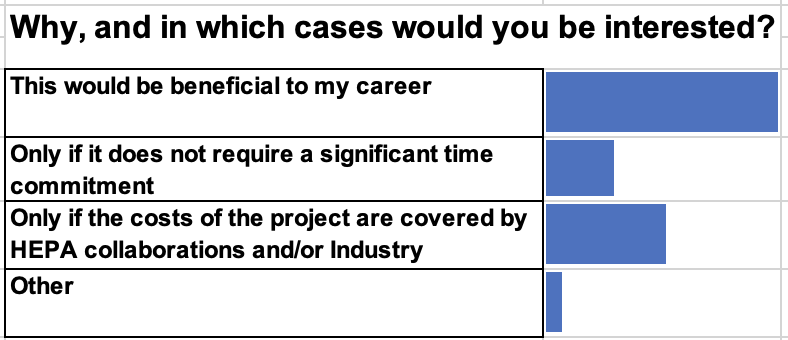}
\caption{}
\label{fig:nonhep_Q34}
\end{minipage}
\end{figure}

Feedback from HEPA alumni provides a sense of the challenges in transitioning from HEPA, and the measures we can take to fix the current shortcomings. Figure~\ref{fig:nonhep_Q69Q70Q66Q68} (top left) shows that almost half of HEPA alumni attempted to find direct employment in the field. Failing to stay in the field reflects the shortage of permanent jobs in HEPA and, simultaneously, lucrative opportunities in industry, particularly higher salaries.  Figure~\ref{fig:nonhep_Q69Q70Q66Q68} (bottom right) shows the variety of sectors in industry that alumni currently work in, with a majority of them in STEM-related fields. Figure~\ref{fig:nonhep_Q69Q70Q66Q68} (bottom left) shows the final involvement level with HEPA before exiting the field. Almost 50\% of responding alumni transitioned after the completion of Undergraduate, Masters or PhD degrees or after postdoc position(s). Figure~\ref{fig:nonhep_Q69Q70Q66Q68} (top right) shows that the most important resource to obtain industry jobs were networking and co-workers. Figure~\ref{fig:nonHEP_Q71} shows that almost all HEPA skills are valuable in the current industry jobs. Another stated goal in feedback from HEPA alumni was to build synergy between HEPA and its alumni. While it might be challenging for alumni to return to HEPA as indicated in Figure~\ref{fig:nonhep_Q72}, a large fraction are willing to participate in projects connecting HEPA and industry, as shown in Figure~\ref{fig:nonhep_Q33}, and many see value in doing so, as indicated in Figure~\ref{fig:nonhep_Q34}.

\section{Career transition points}
A career in HEPA is highly desired by early career scientists. However, academic jobs are scarce. It is difficult to predict the availabilities and the possibility of obtaining a permanent position in a research institute, lab or  university. HEPA researchers and physics degree holders are often forced to transition to non-HEPA careers at different career stages and due to various reasons, as is evident from  Figure~\ref{fig:nonhep_Q69Q70Q66Q68} (bottom left). Career decisions are guided by long-term priorities in life; such as, e.g., aspirations, interests, values, economic situation, financial security and personal situations among many others. Reasons to pursue a degree or particular area of research may evolve over the course of a degree at the undergraduate, graduate (Masters/PhD), postdoc or even more senior levels. On the other hand, physics degree holders typically have numerous job and career opportunities available that apply their skills outside the field. Indeed, new job openings are continually driven by factors such as e.g. technological advances, automation of manufacturing industries, IT and internet revolution, environmental challenges requiring soil conservation and precision agriculture, to name a few examples. Hence, physics degree holders have a greater than ever opportunity to build a career in industry, and the challenge is building bridges between those  seeking jobs --- for whatever reason and whatever level of seniority ---  and careers in non-traditional HEPA areas. References~\cite{nonhep_aipbachelors} and~\cite{nonhep_spscareer} list, by U.S. state, the employers and fields of Physics bachelors and their employment sectors, respectively. Reference~\cite{nonhep_flowchartcareer} shows a flowchart of possible career paths for Physics students with Bachelors, Masters and PhD degrees. Figure~\ref{fig:nonhep_Q42} from our Snowmass survey shows that around half of people in our community consider leaving academia and that half of those (Figure~\ref{fig:nonhep_Q69Q70Q66Q68} top left)  who leave academia try to obtain a job in academia prior to exiting the traditional field.

Students with an Undergrad or Masters degree in physics may transition directly to non-research teaching or industry jobs, or to a different field of specialisation, and then from there move on to an academic or industry position. At this level of transition, research may not be a significant experience for the majority. Hence, a HEPA and/or non-HEPA physics research skill set would not be a significant part of their CV or resume, which typically would contain only educational attainments in the core physics degree program. Physics students getting jobs after Undergrad or Master levels might not be committed to pursuing research due to economic, financial and/or personal reasons as contributing factors. Even if  Masters alumni  may have some advantage over Undergrads with more research experience, arising from an academic curriculum requiring a thesis in a research topic(s).

A PhD or postdoctoral experience in HEPA or any other physics research area typically features a significant research component. It is also designed to develop critical research, analytical and writing skills, as well as the ability to effectively work both independently and in collaboration with other researchers. This puts those with such experience at a significant higher pedestal of skill set when it comes to seeking jobs in industry or academia. The reasons for PhD students and postdocs to go for non-HEPA or industry jobs could largely be the same as for students who exit after an Undergraduate or a Masters degree, but those decisions may not be very easy to take after investing years into research,especially after a postdoc. 

During the course of working in the Snowmass CPD-WG meetings over the past several months we had presentations and direct discussions with several HEPA-alumni. Most of them exited the field after a PhD degree or postdoc and some of their experiences are presented here~\cite{nonhep_remington,nonhep_vasel,nonhep_pasner}. The most important step was the difficult decision to make up their mind to leave academia at such relatively late stages, which also connects to personal matters for many. A professorship comes with its lifetime of studying science, and with a very strong appeal of a comfortable intellectual life devoted to understanding nature at its most fundamental level. However, the path to a professorship comes with many challenges and uncertainties in terms of work/life balance, pressure to publish, competition with scientists across the globe for a limited number of jobs, limited choice of geographic location, pressure to secure grants, and lesser material compensation compared to the industry. At the same time, many fulfilling career paths exist for physicists who are willing to embrace a transition to industry. For some, what is sought is an industry job that is sufficiently interesting and exciting, so as to allow the use of current skill set in a creative way, as well as  challenging enough to foster the development of new skills and allow contributions to have a positive impact in the world. Quote from the presentations says - \textit{Even though for the industry transition it may be hard to get "out of the box" but the only way to make sense out of "change" is to plunge into it, move with it, and join the "dance"}. This applies to exiting the field at any level. Though rare but not uncommon for HEPA physicists to work with non-profit organisations or as science writers~\cite{nonhep_sciencewriter1,nonhep_sciencewriter2} or other academic and scholarship activities; for example, ~\cite{nonhep_yangyangcheng} when personal passion is a big driving factor. Though even more rare, but even faculty members (tenured or nontenured) may also seek industry jobs for personal or other reasons.

\section{Intersections with other Snowmass groups}
As already mentioned, skills developed in HEPA are in demand in  industry. The Snowmass WGs mentioned below highlight how these skills learnt in experiments are not only valuable for  future of HEPA but also serve as basis for future innovation in HEPA and industry. There are several careers in industry tied to these skills as described here.

\textbf{AF1}: \textit{Beam Physics and Accelerator Education}~\cite{nonhep_af1}: Some of the trained particle physicists go into full time accelerator physics careers that maintains unique capabilities and develop new technologies for accelerators. A worldwide exchange of scientists and young researchers between accelerator institutions can play an invaluable role in training the next generation of accelerator physicists and engineers, who can then work work towards developing accelerators for the future HEP. They can  also benefit society in fulfilling the demand for trained accelerator physicists in  the development of high-quality beams and energy-efficient technologies for medicine. Applications of accelerator science in Medical therapies, photon and particle probes in industry, material science, chemistry, biology, pharmaceutical development and applied nuclear science will continue to stimulate demand for the expertise of well-trained specialists in accelerator science and technology and accelerator related software.\\
\textbf{NF05/10}: \textit{Neutrino properties and Neutrino Detectors}~\cite{nonhep_nf5nf10}: Small scale experiments, like in neutrino or nuclear physics community, can provide a healthy breeding and training ground for careers of PhD and postdoc researchers, by providing a unique set of diverse opportunities. It can serve as a bridge to a gateway of career options in academia or industry. Due to the typical short span of such experiments young members can take ownership of significant aspects of a project throughout their tenure, and productively contribute to multiple aspects like the design, construction, operations, and data analysis. This enables early career scientists from these experiments to enter the job market with a strong and transparent portfolio, in terms of publications and contributions that may tend to get obscure in big experiments, especially when applying for jobs within or outside HEPA.\\
\textbf{IF/AF6 and CommF3}: \textit{Instrumentation Frontier}~\cite{nonhep_if2}, \textit{Advanced Accelerator Concepts}~\cite{nonhep_af6} and \textit{Machine Learning}~\cite{nonhep_commf3}: Training in Detector and Instrumentation Technologies and engagement with the industry in applications and technology transfers from the particle physics community can greatly enhance career transition to a related industry. Schools like  ISOTDAQ~\cite{nonhep_isotdaq}, EDIT~\cite{nonhep_edit} and ESHEP~\cite{nonhep_eshep} and the CERN/FNAL Collider Schools~\cite{nonhep_colliderschool} can stimulate young careers that can fuel innovation not only in HEPA but also in  industry. Application of Machine Learning in all aspect of HEPA wherever vast data and its pattern need to be understood to discover physics or monitor data quality from detector and instrumentation has a great demand in data science in industry.

\section{Existing programs}
Many resources currently exist for finding industry jobs, and resuming building and identifying career paths for physics degree holders; e.g.,~\cite{nonhep_aipcareer,nonhep_apscareerprivate,nonhep_apscareerservice,nonhep_physicstodayjobs}. However, one element that is missing is a platform for networking. As Figure~\ref{fig:nonhep_Q69Q70Q66Q68} (top right) indicates, networking and coworkers constitute the most significant factor when it comes to finding industry jobs. For the past four years, networking events for industry careers have been regularly organised by CERN and have proven to be a big draw.  It is obvious that this is one area where we need to focus more. CERN conducted a survey in 2020 within its alumni community concerning their expectations of the Alumni Network. It confirmed that early career community in HEPA have little to no support for moving out of academia and the idea of working in industry is completely alien to them. Subsequently, the CERN office for alumni relations\cite{nonhep_networkCERNalumni} discovered an abundance of goodwill and enthusiasm on the part of CERN alumni to 'give back' and share their experience with a 'younger version of themselves'. A candid conversation, highlighting career trajectories out of academia to other sectors and focusing on challenges and hurdles through the interviews, has been much appreciated by the soon-to-be CERN alumni. It is worth pointing out that the initial seeds of CERN alumni were sown by the CMS Collaboration~\cite{nonhep_networkCMS2013,nonhep_networkCERN2013,nonhep_cmslinkedin} around 2013 and, given its subsequent success, it was adopted by all CERN  Experiments~\cite{nonhep_networkCERN2016,nonhep_networkCERN} and grew into the official CERN alumni network.  The CERN Alumni program now has over 7500 active members, 85\% of whom are alumni or soon-to-be alumni. Having an annual budget of 35 - 50K Swiss Franks (most of which is allocated for the alumni.cern platform). To achieve one of its goals to support those in their early careers,  it focuses on the following:

\begin{itemize}
\item Experience sharing by alumni  on moving out of Academia at events organised by CERN,  spotlights on CERN Alumni focuses on career transitions and experience. 
\item Jobs board where alumni and companies post job opportunities. The Office for Alumni Relations piggybacks on HR career fairs, where companies can hope to recruit from within the CERN alumni talent pool. 
\item Virtual Company showrooms  where companies having posted job opportunities can take part in 1 hour events, 30 mins presentation and 30 mins Q \& A sessions.
\item Mentoring module, giving a short video summary of HEPA alumni experience in getting an industry job.
\end{itemize}
The Fermilab Student and Postdoc Association (FSPA) ~\cite{nonhep_fspafermilab}, among other activities, organises Career Panel discussions and talks regularly. It also has a linkedin group for networking with its alumni. In this context, one can exchange job information, share interesting ideas, and build ones own network. FSPA represents the community of early career researchers at Fermilab, including all students, postdocs, summer students, and interns.

The HEP national labs represent a quite wide community, encompassing thousands of users from hundreds of universities and several countries. As demonstrated above, it can serve as a very strong hub for career networking.  We do not discuss here any existing career programs at academic institutional/universities, such as job fairs or targeted campus visits by business/trade group, as they are of more general nature. Nonetheless these are still useful to the community members who happen to be at those institutions. 

\section{Methods to strengthen existing programs}

Getting a tenure track or a scientist position or a job in industry is a process that needs right preparation. Being in HEPA one may be easily aware of the process of getting a tenure track or a scientist position. It is also is easy to find mentors who can be helpful for that purpose, being surrounded by established scientists, including one's own supervisor. However, when it comes to an industry job in STEM or otherwise, it is an uncharted territory for many, as HEPA alumni who move to industry are not accessible for guidance due to lack of organised networking or communications. In fact, HEPA alumni rarely have channels or motivation to connect with their alma mater. 

The existing programs, though few, can be broadly divided into networking and skill building. HEPA and industry relationship has always been synergistic where sharing and benefiting from technological advances and know-how pushes the envelope of knowledge. However, when it comes to building relationships with industry in terms of networking, transition to industry jobs and collaborations with alumni on projects of mutual interest, the realisation and progress is just beginning. It is clear that there are very few existing programs like at CERN that are fairly mature in establishing networking structure for industry jobs. Elsewhere, the efforts are nascent to non-existent and need to be developed and strengthened.  One huge advantage in pursuing these efforts at the national labs is that they serve as the hub of HEPA community and can have structural as well possible financial resources for this common good. Clearly, there is also a role for U.S. funding agencies to support efforts at career-building through targeted supplemental funding. The survey results clearly show a need and desire to do so. Seeking funding to support and strengthen HEPA-industry alumni relations would be a big boost to early career transition to the industry.

HEP labs technology public-private partnerships  \cite{nonhep_fnalindustry,nonhep_cernindustry} with industry exist in Accelerator Technologies, Computers Information Science,
Detector and Engineering Technologies and also Environment Safety. This can be leveraged to create experience for resident students and early career scientists and build connections for a future industry career. Internships in instrumentation and accelerator know-how can strengthen careers in medical science and industry. As an example, partnership of University College London’s Centre for Doctoral Training in Data Intensive Science~\cite{nonhep_uclindustry}  with the high-tech computing sector is greatly facilitating the transition to employment in this sector for those who desire it. There, even HEP and cosmology theorists get internships in industry during their PhD. Such internships can be beneficial for enhancing hiring pools at HEP labs and within particular industries.

Besides networking and skill building, there are several other factors that determine finding jobs in industry and those need to be strengthened too. These are described in the following section.

\section{New ideas, new partners}

HEPA young careers, while pursuing research, acquire several skills like computing, software, detector design, logical thinking, publishing, presenting professional work at conferences, statistical analysis, simulation, and hardware development. This makes them highly employable within and outside academia. Such skills can also serve as attractors for students who may not want to pursue a PhD in HEP  or a career in acedemia but use them as a springboard to other STEM careers in industry. This makes it very important to expand our nascent ongoing efforts to engage early career HEPA scientists with more senior scientists and facilitate the process of transition to industry jobs. We must  scale up and  have a model to sustain these programs. The experience, survey feedback and efforts described above strongly indicate that HEPA alumni are more than willing and happy to respond and engage. However, within HEPA we need to be more motivated and creative on how to realise this engagement.

\subsection{Engagement of HEPA supervisors and mentors} Attempting to transition from a completely academic work role to an industry job is a daunting task. It is not a well-defined path and one needs to take initiative and explore many options. First and foremost is the direct involvement of HEPA supervisors and mentors in the process. This is important as HEPA supervisors may not themselves be knowledgeable to advise or mentor to help pave paths towards industry jobs, and a reluctance to be involved in the process may also arise from research commitments and related priorities. But their involvement will create comfortable and supportive environment. They must be sensitised and trained to guide and be be proactive to support  early career scientist's participation in opportunities that prepare for careers outside HEPA.  They must allow a certain fraction of time to pursue opportunities and preparation activities for a possible industry career. Indeed, there is overwhelming desire for this, as shown in Figure~\ref{fig:nonhep_Q62}, Workshops training supervisors to understand the needs of their mentees and ways to improve methods for job preparation are thus called for.

\subsection{Networking}The value of networking and making connections is always high for any job search, and all the more so for industry jobs.  Highly sought after skills like computing, software, detector design, logical thinking, statistical analysis, simulation, and hardware development are not the only factors in this process. Networking is an additional personal skill that contributes to social well-being, exchange of ideas, meeting people at professional levels and boosting one's professional confidence. The best career opportunities often come from the connections such as former classmates and co-workers. Nowadays, companies are increasingly looking towards universities for fresh talent, and are eager to meet students who are looking for a career in industry. In addition, reaching out to professors, alumni and other students to get advice and information about career opportunities puts one at an advantage. We need to develop technical tools to improve networking. Tools like \cite{nonhep_networkCERNalumni} are required to serve our community of early career and alumni for jobs as well as mutual connections. Just like CERN experiments, US based HEP community is also very big, varied and interlinked; for example, in neutrino  and dark matter detections experimental collaborations, and links to astrophysics and related fields like nuclear and gravitational physics, with common learnt skills. 
Hosting a dedicated platform in the US to connect and engage HEPA with its alumni is required, just like at CERN \cite{nonhep_networkCERNalumni}. This would not only enable an organised framework, and serve as a hub to facilitate this process, 
but several other actions in this direction can be planned and executed on that basis.
A DOE lab would be an ideal place to host it, like Fermilab, which is a hub for  US particle physics. However, this would need a small amount of investment to create a database of alumni and need technical and personnel support to develop and maintain it.

\subsection{Job search process} Many fulfilling career paths exist for physicists who are willing to prepare for, and embrace, a transition to industry. However, translating HEPA abilities and adapting skills to an industry job can require special preparation and many find themselves unfamiliar with this process. Preparing an industry tailored resume is one such important thing. For example, very few industry job requisitions come with “PhD Physicist” requirement in the title, and one should consider roles with Engineer, Data Scientist, Algorithm Developer, Quantitative Analyst, Research Scientist in the title, and convert  CV to a resume suitable for each target industry like finance, defense, tech, software etc. Focusing on skills that generalize and core competencies that are relevant to the job search are of far more use;  physics jargon tied to previous research, like  “Searched for New Physics or Higgs Properties”, should be strictly avoided. Emphasize research publications via a link to SPIRES (not enumerating scores of publications and articles), ability to work in multi-disciplinary team environment and highlighting any distinctions, awards, major accomplishments in generic terms is essential to be understood by industry hiring panel. Careful attention must be paid to studying the company profile, past hires, culture of embracing STEM graduates and PhDs and chances of career advancement. HEPA alumni can be a big boon in helping out with this preparation, career advice and sharing their own experience. 

\subsection{Projects and Internships} While there are many examples of internships sponsored by the National Labs and funding agencies, like DOE and NSF, at the labs or universities, there are significantly fewer internship opportunities for HEPA early career scientists or students in industry. Industry internships can serve as a very good opportunity to experience industry culture and environment and apply HEPA skills. The biggest impediment for these internships would be a time off from core research to allow a period of time to focus on industry internship. We must look for mechanism to integrate this in core research curriculum. Most HEPA research is funded by federal grants and having time-off from research would require change in funding rules and regulations and  above all, change in mindset of supervisors. While industry internships for students in general exist, internships for more senior researchers at PhD or postdoc level may not be desired, or hard to fund, by  industry. However, naturally, any industry experience would be very valuable for industry job and career.

Tailoring an academic  experience  for  a  position  in  industry  will  help  prepare early career HEP scientists for industry jobs. While industry internships funded by  industry may be harder at PhD or postdoc level, it would be  easier for HEPA to create its own focused opportunities at all levels to train its people for industry jobs. While skills like software are in common use in HEPA, how they are used and applied in a goal and profit oriented industry setting, with time bound targets and deliverables, requires dedicated training opportunity. Job prospects in the medical field, like working with particle accelerators for radiation therapy for cancer treatment and similar applications, might benefit from dedicated experienced with experimental apparatus like particle detectors and accelerators. Working with detectors leads to a rich experience in electronics and instrumentation, but the duration of time spent may be short as the end goal is always to produce physics results which take significant time of a researcher and  not relevant for an industry job. Training for industry jobs by HEPA  needs careful planning and should not dilute the focus on core research goals, but at the same time it should prepare those who wish to transition and are willing to go extra mile to find a proper opportunity.

\subsection{Collaborations}
Alumni are relatively a low cost but a very valuable asset with an abundance of experience of transitioning from HEPA. Their goodwill to contribute and strengthen ties with HEPA can be tapped to facilitate industry job transition.  Figure~\ref{fig:nonhep_Q33} and Figure~\ref{fig:nonhep_Q34} indicate that HEPA alumni are interested in participating in projects that connect HEPA with industry in case it is beneficial to their careers as well. Hence, one attractive way to encourage and engage alumni is to be proactive and creative in collaborating with them on HEPA projects. While several private companies and enterprises may have legal restrictions against allowing individual level alumni participation, in cases where it is allowed we must provide opportunities to collaborate. In this way alumni will feel welcome to their alma mater in a more constructive way than to be just industry job facilitators. This individual collaboration can be extended to the company of the alumni itself and this can strengthen  knowledge transfer from labs and universities and vice versa; and Work done by HEPA research can benefit companies and vice versa. A recent example has been the use of Amazon Web Service (AWS) with the CMS experiment workload through HEP Cloud project~\cite{nonhep_aws}. HEP experiments require massive computing resources in irregular cycles and using commercial clouds is an important ingredient in this, where resources can expand or contract on demand.These collaborations might also lead to first steps towards reversing the brain drain from industry to HEPA. Survey result Figure~\ref{fig:nonhep_Q72} indicates a significant interest in alumni to return full time to HEPA, especially if the compensation match as well as work part-time. 

\section{Conclusions}
A majority of HEPA early career scientists will inevitably transition out of traditional HEPA careers, due simply to the structural disparity between the relatively few traditional HEPA career positions available in any given year and the much larger number of available HEPA early career scientists desiring to fill those positions. Personal factors, such as the larger salaries potentially available in industry and negative factors such as burnout, can also contribute to the desire of early career scientists in transitioning out of traditional HEPA roles in academia and the national laboratories. Whatever the reason, the HEPA community has both a pragmatic need and an ethical obligation to connect early career HEPA job seekers with the resources and relationships which will allow them to identify interesting careers and obtain stable positions that will allow them to advance professionally as well as personally. 

Each major milestone along the path from undergraduate to post-doctoral researcher represents a potential transition point where a young HEPA physicist would ideally be empowered with the information they need to develop an informed and realistic perspective on the full breadth of opportunity available to them, within both traditional HEPA as well as non-traditional careers. Many HEPA students and young scientists arguably continue to pursue an entirely academic career path out of inertia and lack of awareness of the broader world of opportunities avaiable outside of traditional HEP because of the relative dearth of resources connecting non-traditional HEPA industries and employers to the pool of highly skilled HEPA physicists entering the job market in any given year.

Above, we have identified, through surveys of early career HEPA scientists, networking and mentoring relationships with earlier generations of HEPA physicists as the single most effective tool in communicating career opportunities to junior scientists. Very clearly and self-evidently, this is a highly effective strategy given that nearly every physics student, when approaching a career transition point such as graduation or the end of a post-doctoral term, seeks to emulate their direct supervisor and immediate colleagues by tenaciously pursuing a traditional HEPA career to the exclusion of all else. We have shown above that it is possible to break out of this self-reinforcing dynamic through a variety of mechanisms which can nurture and develop bonds and help establish more broad communities of interest between mid- and late-career HEPA scientists in non-traditional careers than has been the case throughout much of the history of HEPA in the United States. Mechanisms such as e.g. alumni networks and opportunities for internships in non-traditional HEPA settings have achieved great success in structurally connecting early career job seekers with both the information and the opportunities which they need in order to effectively pursue a career that is ultimately both professionally and personally rewarding. In addition, private employers reap the significant economic benefits of employing HEP physicists skilled in both STEM disciplines and related skills such as technical writing and collaborative working, while the wider society benefits from the diffusion of highly talented physics professionals from academia into the overall economy, strengthening both society at large as well as improving quality of life for each individual.

While organizations such as the APS and national lab user groups can tailor and offer programs that meet their specific goals, it is clear that a more broad and systematic approach to the marketing of career opportunities to early career physicists is needed beyond the efforts of such organizations if optimal outcomes at both the structural and personal levels are to be achieved for HEP physics as well as the wider society. However, establishing and institutionalizing these approaches, as well as socializing them with members of the HEPA community from the most junior to the most senior levels, will require both dedicated effort from the community along with relatively modest funding in order to bring early and mid-to-late career scientists together in environments and settings conducive to formal and informal networking and the building of relationships that will last a lifetime.

\section*{References}
\bibliography{iopart-num}


\end{document}